\documentclass[prl,aps,epsfig,showpacs,notitlepage]{revtex4-1}
\usepackage{amsmath,amssymb,amsfonts,amstext}
\usepackage{graphicx,breakurl,color,setspace}
\usepackage[T1]{fontenc}
\usepackage[latin9]{inputenc}
\usepackage[normalem]{ulem}
\usepackage[unicode=true,pdfusetitle,pdfborder={0 0 1}]{hyperref}
\usepackage{upgreek}
\usepackage{bm}
\usepackage{mathrsfs}
\usepackage[symbol]{footmisc}

\hypersetup{colorlinks,
linkcolor=blue,          citecolor=blue,        filecolor=blue,      urlcolor=blue           }

\newcommand{\ket}[1]{| #1 \rangle}

\graphicspath{{../figures/}}

\begin{document}

\title{Observation of topological links associated with Hopf insulators
\\in a solid-state quantum simulator}
\author{X.-X. Yuan$^{1\dagger}$\footnote[0]{$^{\dagger}$
These authors contributed equally to this work.}, L. He$^{1\dagger}$, S.-T. Wang$^{1,2\dagger}$\thanks{s},
D.-L. Deng$^{1,2,3}$, F. Wang$^1$, W.-Q. Lian$^1$, X. Wang$^1$, C.-H. Zhang$^1$, H,-L.
Zhang$^1$, X.-Y. Chang$^1$, L.-M. Duan$^{1,2*}$\footnote[0]{$^{*}$
Corresponding author. Email: lmduan@umich.edu}}

\affiliation{$^1$Center for Quantum Information, Institute for Interdisciplinary Information Sciences, Tsinghua University, Beijing 100084,
PR China} \affiliation{$^2$Department of Physics, University of Michigan,
Ann Arbor, Michigan 48109, USA}
\affiliation{$^3$Condensed Matter Theory Center and Joint Quantum Institute, Department
of Physics, University of Maryland, College Park, MD 20742-4111, USA}

\begin{abstract}
Hopf insulators are intriguing three-dimensional topological insulators characterized by an integer topological invariant.
They originate from the mathematical theory of Hopf fibration and epitomize the deep connection between knot theory and
topological phases of matter, which distinguishes them from other classes of topological insulators. Here, we implement
a model Hamiltonian for Hopf insulators in a solid-state quantum simulator and report the first experimental observation of their
topological properties, including fascinating topological links associated with the Hopf fibration and the
integer-valued topological invariant obtained from a direct tomographic measurement. Our observation of topological links
and Hopf fibration in a quantum simulator opens the door to probe rich topological properties of Hopf insulators in
experiments. The quantum simulation and probing methods are also applicable
to the study of other intricate three-dimensional topological model Hamiltonians.
\end{abstract}

\maketitle

\section{Introduction}
The interplay of topology and symmetry plays a key role in the
classification of quantum phases of matter \cite{hasan2010colloquium,
Qi:2011wt, Kitaev2009Periodic, Schnyder:2008ez, Chen2012symmetry}. It gives
rise to a notable periodic table for topological insulators and
superconductors for free fermions \cite{Kitaev2009Periodic, Schnyder:2008ez}%
. The recently discovered $\mathbb{Z}_{2}$ topological insulators \cite%
{konig2007quantum, hsieh2008topological, Chen2009Experimental}
fit into this classification paradigm.
Topological insulators are insulating materials featuring
conducting surface states protected by time-reversal symmetries \cite%
{hasan2010colloquium, Qi:2011wt}. Without any
symmetry protection, multi-band Hamiltonians in 3D should only exhibit
topologically trivial phases according to the periodic table \cite%
{Kitaev2009Periodic, Schnyder:2008ez}.

With broken time-reversal symmetry, Hopf
insulators \cite{moore2008topological} are intriguing three-dimensional (3D)
topological insulators that elude the standard classification paradigm of
topological phases for free fermions \cite{Kitaev2009Periodic,
Schnyder:2008ez}. Recently, Moore, Ran and Wen~\cite%
{moore2008topological} realized that the two-band case is special because of
the existence of Hopf map, a topological map linking the 3D torus that represents the momentum space
with the 2D Bloch sphere that describes the state space of a two-band Hamiltonian. Hence,
the topological Hopf insulators could exist for two-band Hamiltonians in 3D space, which lie outside of the standard classification paradigm of
topological phases. Implementation of Hopf insulators, however, poses a formidable experimental challenge \cite%
{Deng2013Hopf, Deng2016Probe}. Recently, quantum simulation platforms have
proven to be well suited for the experimental study of 1D and 2D topological
insulators \cite{Roushan2014Observation, Schroer2014Measuring,
Kong2016Direct}. We extend this tool to study more intricate 3D topological
models that have not been realized yet in any other experimental platform.

In this paper, we implement Hopf insulator model Hamiltonian in a solid-state quantum simulator
and report the first experimental observation of its topological properties, including
the topological links associated with the Hopf fibration and the integer-valued topological invariant.  Our quantum simulator
is realized with the nitrogen-vacancy (NV) center in a diamond sample. The diamond NV center has recently emerged as a
promising experimental system for realization of quantum computing,
simulation, and precision measurements \cite{Gruber1997Scanning,
Doherty2013The, Cai2013ALarge, Childress2013Diamond, Childress2014Atom}. The
key observation here is that the Hamiltonian for free fermions is
diagonal in the momentum space, so there is no entanglement between
different momentum components in its ground state. We can measure the
quantum states for each momentum component separately in experiments \cite%
{Roushan2014Observation, Deng2014Direct}, and the properties of the whole
system can then be determined by collating individual measurements. We measure the
integer-valued topological invariant by a direct 3D integration over the parametric
momentum space \cite{Deng2014Direct}. Through quantum state tomography, we
experimentally demonstrate the Hopf fibration with fascinating topological
links, showing clear signals of topological phase transitions for the
underlying Hamiltonian.

\section{Results}
\subsection{Quantum simulation of Hopf insulator model Hamiltonian}

\begin{figure}[t]
\includegraphics[trim=0cm 0cm 0cm 0cm, clip,width=\linewidth]{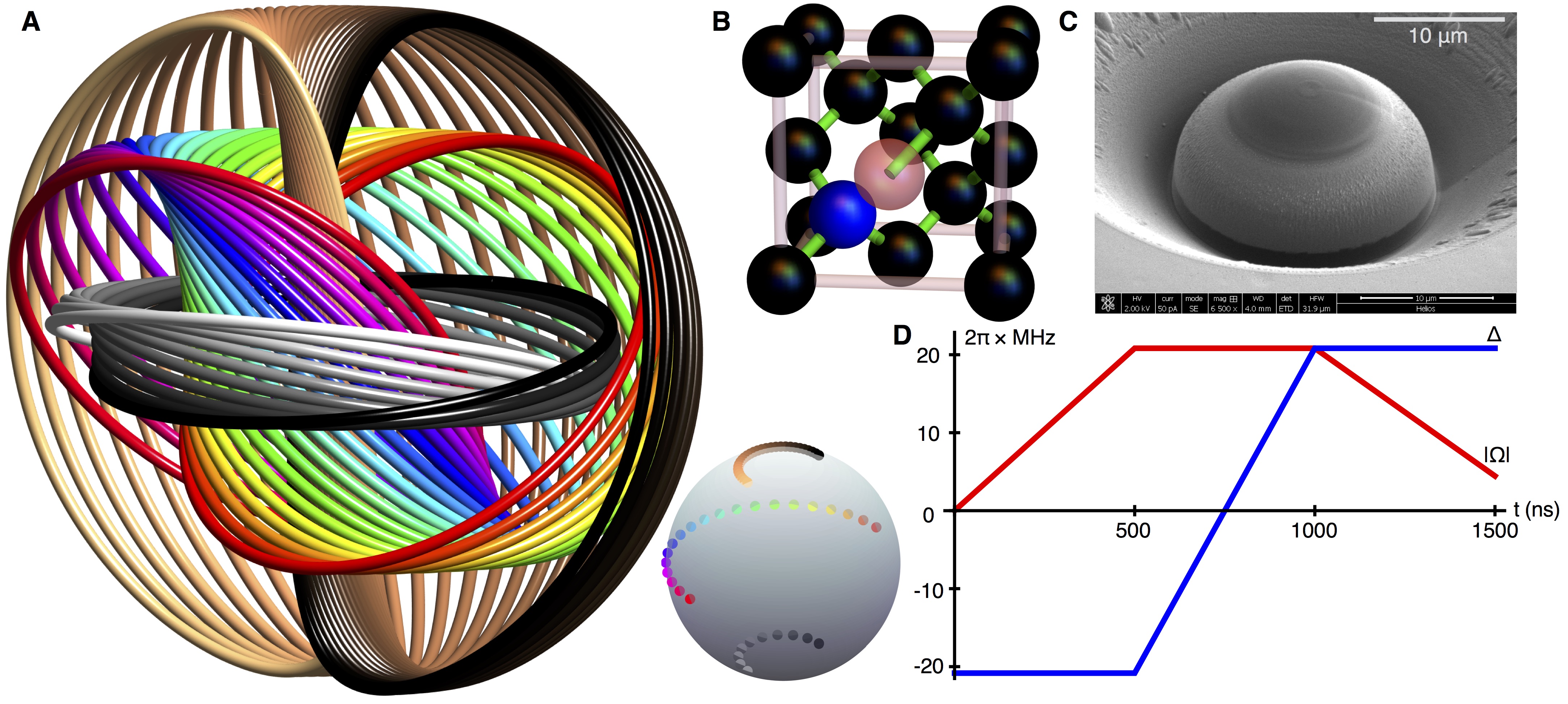}
\caption{\textbf{Illustration of Hopf fibration and the experimental system.}
(\textbf{A}), Hopf fibration, where each point (distinguished by color) on the
Bloch sphere in the right corner is mapped to a closed loop in the 3D space
by the Hopf map and all loops are pairwise linked to each other (see
Supplementary Information for the explicit form of the Hopf map in our
system). (\textbf{B}), A diamond nitrogen-vacancy center with its defect spin
used for the quantum simulation of the Hopf insulator. (\textbf{C}), A
solid-state immersion micro-lens is fabricated on top of the diamond
nitrogen-vacancy center to increase its detection efficiency by optical
readout. (\textbf{D}), A typical path of adiabatic passage for the parameters
in the Hamiltonian. The example shown here is for adiabatic preparation of
the ground state of $H_{\mathbf{k}}$ with the momentum components $(k_{x},
k_{y}, k_{z})/2\protect\pi = (0.4,0.3,0.5)$ and the parameter $h=2$.}
\label{Fig:Fig1}
\end{figure}

A general two-band Hamiltonian in the momentum space can be written in the form
\begin{equation}
H=\sum\limits_{\mathbf{k}}\Psi _{\mathbf{k}}^{\dagger }H_{\mathbf{k}}\Psi _{%
\mathbf{k}}=\sum\limits_{\mathbf{k}}\,\Psi _{\mathbf{k}}^{\dagger }\hbar
\Omega \mathbf{u}(\mathbf{k})\cdot \mathbf{\sigma }\Psi _{\mathbf{k}},
\end{equation}%
where $\hbar \Omega $ denotes the energy unit, the dimensionless
coefficients $\mathbf{u}(\mathbf{k})=(u_{x},u_{y},u_{z})$ represent certain
functions of 3D momenta $\mathbf{k=}\left( k_{x},k_{y},k_{z}\right) $, $\Psi
_{\mathbf{k}}=(a_{\mathbf{k}\uparrow },a_{\mathbf{k}\downarrow })^{\intercal
}$ are fermionic annihilation operators with pseudo-spin states $|{\uparrow }%
\rangle $ and $|{\downarrow }\rangle $ at momentum point $\mathbf{k}$, and $%
\mathbf{\sigma }=\left( \sigma _{x},\sigma _{y},\sigma _{z}\right) $ are
Pauli matrices. The (composite) Hopf map is a projection from the momentum
space (the Brillouin zone) described by the 3D torus $\mathbb{T}^{3}$ to the
Bloch sphere $\mathbb{S}^{2}$ for the spin-$1/2$ state. For the Hopf map,
the pre-image of each point on the sphere $\mathbb{S}^{2}$ corresponds to a
closed loop in the torus $\mathbb{T}^{3}$---all these loops are
topologically linked to each other, forming a fascinating topological
structure called the Hopf fibration as shown in Fig.~\ref{Fig:Fig1}A. Armed
with the Hopf map, Hopf insulators are characterized by an integer $\mathbb{Z%
}$ rather than a $\mathbb{Z}_{2}$ topological invariant \cite%
{moore2008topological, Deng2013Hopf}. Several two-band model Hamiltonians
supporting Hopf insulators have been constructed in Refs.~\cite%
{moore2008topological, Deng2013Hopf}, and all of them involve complicated
spin-orbital interactions. We consider a primitive model with the
coefficients $\mathbf{u}(\mathbf{k})$ given by
\begin{align}
u_{x}& =2\left[ \sin k_{x}\sin k_{z}+C(\mathbf{k})\sin k_{y}\right] ,  \notag
\\
u_{y}& =2\left[ C(\mathbf{k})\sin k_{x}-\sin k_{y}\sin k_{z}\right] , \\
u_{z}& =\sin ^{2}k_{x}+\sin ^{2}k_{y}-\sin ^{2}k_{z}-\left[ C(\mathbf{k})%
\right] ^{2},  \notag
\end{align}%
where $C(\mathbf{k})\equiv \cos k_{x}+\cos k_{y}+\cos k_{z}+h$ with $h$
being a dimensionless parameter. This Hamiltonian features two distinct
topologically nontrivial phases with the parameter $|h|<1$ and $1<|h|<3$,
respectively \cite{Deng2013Hopf}. Implementation of the Hamiltonian with
cold atoms requires engineering of next-nearest-neighbor spin-orbital
couplings \cite{Deng2016Probe}, which is particularly challenging for
experiments.

Here, we experimentally implement this Hamiltonian and probe its topological
properties with a solid-state quantum simulator represented by a single
nitrogen-vacancy (NV) center in a diamond sample at room temperature shown
in Fig.~\ref{Fig:Fig1}B. The Hamiltonian (1) is
diagonal in the momentum space. To probe its ground state property, we can measure the
quantum state for each momentum component separately in experiments \cite%
{Roushan2014Observation, Deng2014Direct}, and the properties of the whole
system can then be determined by those individual measurements on all momentum components. For each
momentum component $\mathbf{k}$, the ground state of $H_{\mathbf{k}}$ corresponding to the lower band of the Hamiltonian (1) can
be probed through an adiabatic passage in a two-level system, which is
realized through microwave manipulation of the spin levels $\left\vert
0\right\rangle $ and $\left\vert -1\right\rangle $ of a single NV center.
The viability of such an adiabatic procedure is guaranteed by the most
salient feature of a topological phase, the topological gap---as long as the
gap is maintained, topological properties are insensitive to perturbations
or stretching of the energy bands.

We use a NV center under a micro-fabricated solid immersion lens (SIL) to
implement the Hamiltonian $H_{\mathbf{k}}$  for each momentum component $%
\mathbf{k}$. The SIL\ is used to enhance the data collection rate as we need
to scan over many different momentum components to measure the topological
properties of the Hamiltonian. By applying a microwave with phase $\varphi $
to the transition $\left\vert 0\right\rangle \rightarrow \left\vert
-1\right\rangle $, we realize the following Hamiltonian in the rotating frame%
\begin{equation}
H_{u}=\hbar \left\vert \Omega \right\vert \left( \sigma _{x} \cos \varphi
+\sigma _{y}\sin \varphi \right) +\hbar \Delta \sigma _{z},
\end{equation}%
where $\left\vert \Omega \right\vert $ denotes the Rabi frequency of the
microwave and $\Delta $ is the frequency detuning of the spin transition
relative to the microwave frequency. Comparing with the Hamiltonian (1), we
have $\Omega \mathbf{u}(\mathbf{k}) =\left( \left\vert \Omega \right\vert
\cos \varphi ,\left\vert \Omega \right\vert \sin \varphi ,\Delta \right) $.
At the initial time, we take $\left\vert \Omega \right\vert =0$ and prepare
the spin in state $\left\vert 0\right\rangle $, which is the ground state of
$H_{u}\left( t=0\right)$. We then adiabatically tune the microwave Rabi
frequency $\left\vert \Omega \right\vert $ with phase $\tan \varphi
=u_{y}\left( \mathbf{k}\right) /u_{x}\left( \mathbf{k}\right) $ and the
detuning $\Delta $ so that for the final state we have $H_{u}\left( t\right)
=H_{\mathbf{k}}$ in the Hamiltonian (1) for a certain momentum component $%
\mathbf{k}$. A typical adiabatic passage for the parameters is shown in Fig.~%
\ref{Fig:Fig1}D. By this adiabatic passage, we realize the ground state of $%
H_{\mathbf{k}}$, and through quantum state tomography (QST) measurements, we
retrieve its full information. We scan all momentum components $\mathbf{k}$
via the above preparation and detection method to probe the properties of
the full Hamiltonian (1).

\begin{figure}[t]
\includegraphics[trim=0cm 0cm 0cm 0cm, clip,width=\linewidth]{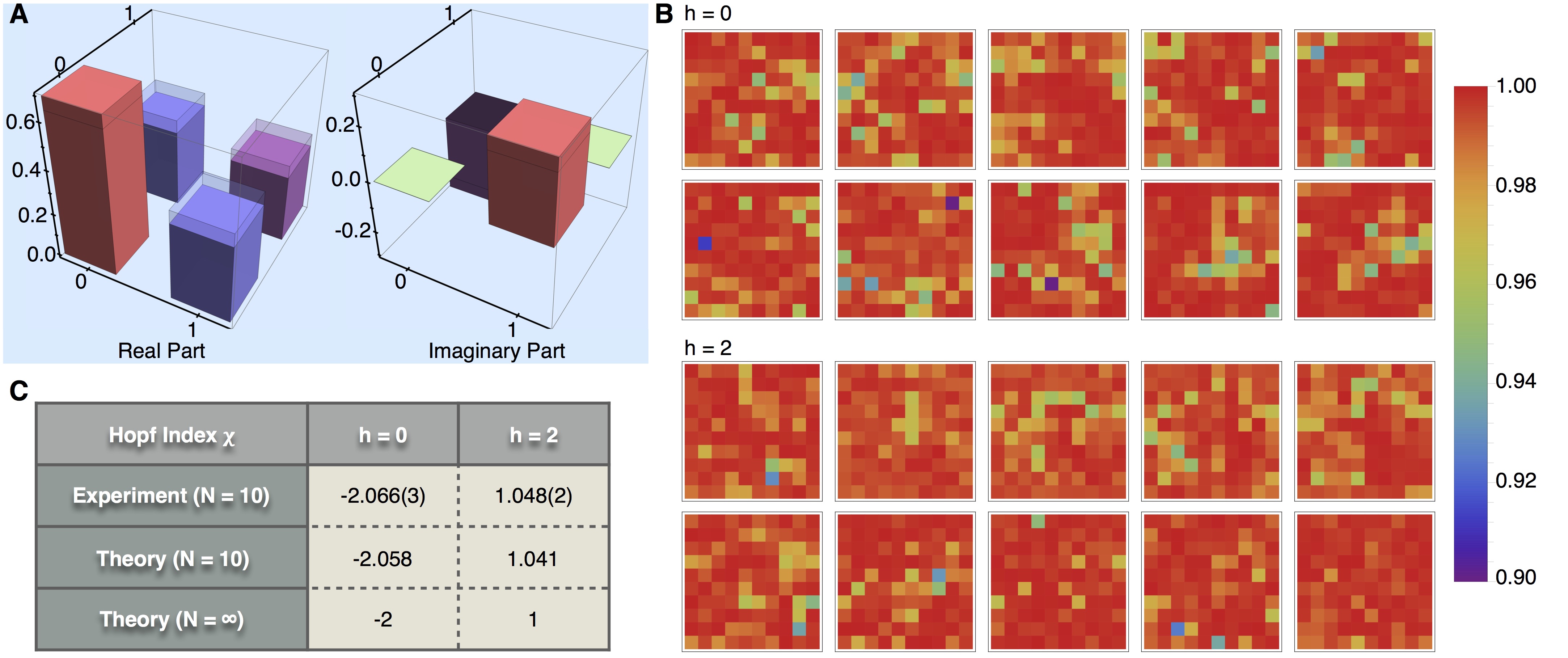}
\caption{\textbf{Measurement of the Hopf index using quantum state
tomography.} (\textbf{A}) Real and imaginary parts of the experimentally
reconstructed density matrix elements. This example shows the ground state
density matrix $\protect\rho _{\mathbf{k}}$ at one particular momentum $%
\mathbf{k}$ with $(k_{x}, k_{y}, k_{z})/2\protect\pi = (0.4,0.2,0.5)$. The
hollow caps correspond to the ideal density matrix elements predicted by
theory. (\textbf{B}) State fidelities $F_{\mathbf{k}}$ measured through
quantum state tomography are shown for different $\mathbf{k}$ with the
values represented by the color map. The upper and the lower panels have the
parameter $h=0$ and $h=2$, respectively. Each panel contains $10$
sub-figures where the momentum $k_z/2\protect\pi$ varies from $0$ to $0.9$
with an equal spacing of $\protect\delta k_z/2\protect\pi=0.1$. The
horizontal and vertical axes of each subfigure denote respectively $k_x/2%
\protect\pi$ and $k_y/2\protect\pi$, which vary from $0$ to $0.9$ with an
equal spacing of $0.1$. The average fidelity for the $10\times 10\times 10$
measured momentum points is $99.1\%$ ($99.2\%$) for the case of $h=0$ ($h=2$%
). (\textbf{C}) The Hopf index from quantum state tomographic measurements
with momentum $\mathbf{k}$ sampled over the $10\times 10\times 10$ mesh. The
number in the bracket represents the standard deviation in the last digit.
The measured Hopf index is close to its ideal integer values for the
corresponding topological phases. The small differences are dominated by the
discretization error of the 3D momentum integration in computing topological
invariants. The scaling of the discretization error with the number of
sampling points is shown in Supplementary Information.}
\label{Fig:Fig2}
\end{figure}

\begin{figure}[t]
\includegraphics[trim=0cm 0cm 0cm 0cm, clip,width=8.9cm]{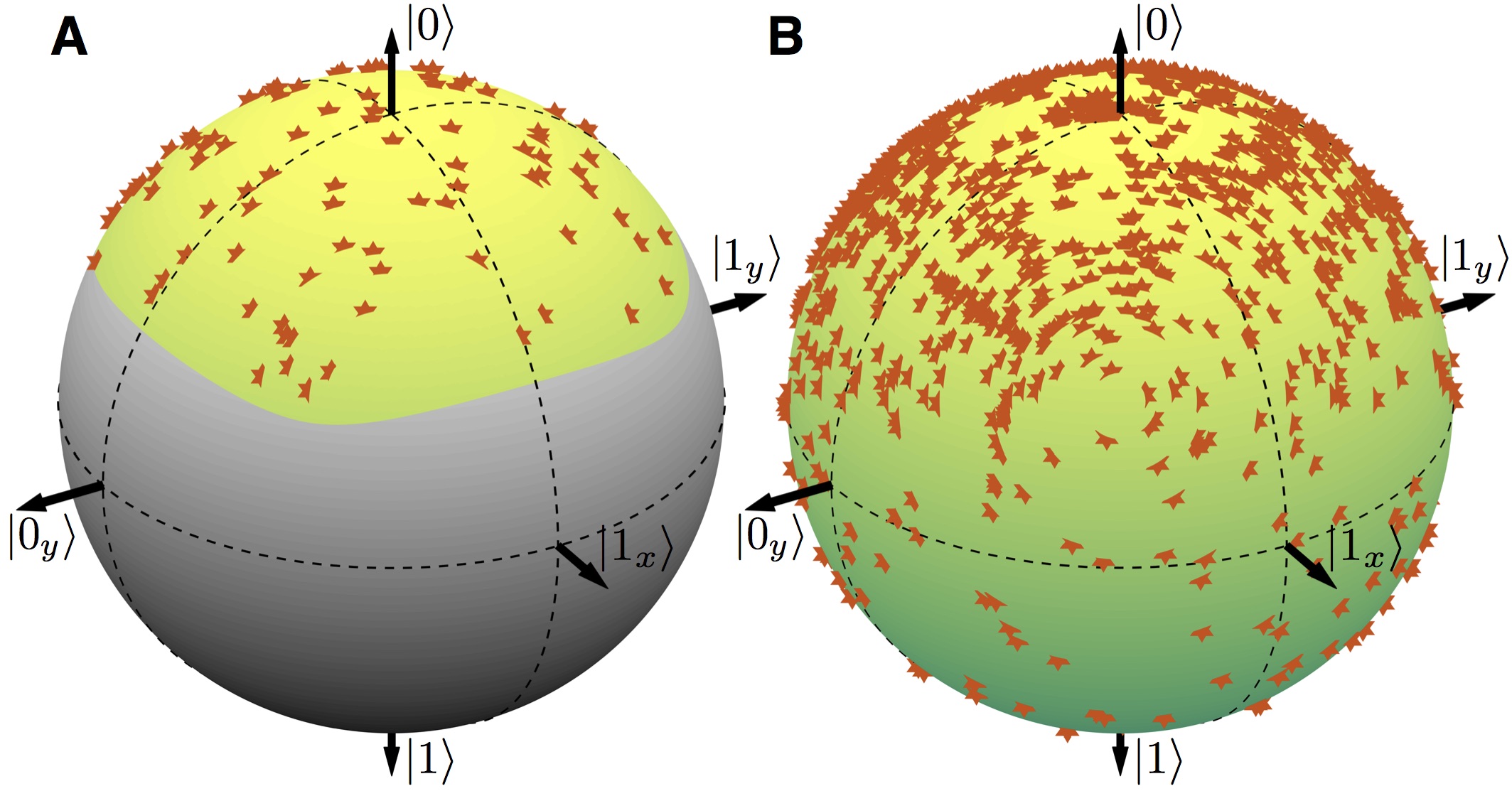}
\caption{\textbf{An intuitive indicator of intrinsic 3D topological
insulator.} (\textbf{A}) The ground-state spin configurations of the
Hamiltonian $H_{\mathbf{k}}$ for $h=2$ are mapped onto the Bloch sphere,
where ${\mathbf{k}}$ is scanned over a 2D cross section with $k_z=0$. The
stars show the experimental data and the yellow cover identifies the
theoretical region from all the momentum points in this cross section. The
partial covering of the Bloch sphere indicates that the Chern number for
this layer is zero. We have verified with our experimental data that Chern
numbers vanish for all 2D cross sections along different directions, so
layered 2D topological (quantum Hall) insulators are not present in our
system. (\textbf{B}) The ground-state spin configurations with ${\mathbf{k}}$
scanned over all 3D momentum points are mapped onto the Bloch sphere. The
Bloch sphere is fully covered, which is consistent with the characteristic
of an intrinsic 3D topological insulator. }
\label{Fig:Fig3}
\end{figure}

\subsection{Measurement of the topological invariant}

To measure the topological properties of the Hamiltonian (1), we use a
topological invariant called the Hopf index \cite%
{moore2008topological,Deng2013Hopf}, which is defined as%
\begin{equation}
\chi =-\int_{ \text{BZ} }\mathbf{F}\cdot \mathbf{A} \, d^{3}k,
\end{equation}%
where $\mathbf{F}$ is the Berry curvature with $F_{\mu }=\left( i/2\pi
\right) \epsilon _{\mu \nu \tau }\left( \partial _{k_{\nu }} \! \left\langle
\Psi _{\mathbf{k}}\right\vert \right) \left( \partial _{k_{\tau }}\!
\left\vert \Psi _{\mathbf{k}}\right\rangle \right) $, $\epsilon _{\mu \nu
\tau }$ is the Levi-Civita symbol with $\mu ,\nu ,\tau \in \left\{
x,y,z\right\} $, $\left\vert \Psi _{\mathbf{k}}\right\rangle $ denotes the
ground state of the Hamiltonian $H_{\mathbf{k}}$, $\mathbf{A}$ is the
associated Berry connection satisfying $\nabla \times \mathbf{A=F}$, and the
integration of $\mathbf{k}$ is over the Brillouin zone (BZ). Depending on
the parameter $h$, the Hopf index $\chi $ takes the following values for the
Hamiltonian (1):
\begin{equation}
\chi =\left\{
\begin{array}{ll}
1, & 1<|h|<3 \\
-2, & |h|<1 \\
0, & |h|>3.%
\end{array}%
\right.
\end{equation}%
So the Hamiltonian $H$ supports two topological Hopf insulator phases and
one topologically trivial phase with the phase boundaries at $h=\pm 1,\pm 3$.

To measure the Hopf index $\chi$, we use the discretization scheme in Ref.~%
\cite{Deng2014Direct}, which provides a general method to probe the
topological invariants in any spatial dimension based on QST in the momentum
space. As shown in the Supplementary Information, the Hopf index $\chi $
quickly converges to its ideal value through mesh sampling over the momentum
space. We sample $\left( k_{x},k_{y},k_{z}\right) $ into a mesh of $10\times
10\times 10$ points with equal spacing and for each $H_{\mathbf{k}}$, we
perform QST to measure its ground state density matrix. A typical
reconstructed density matrix is shown in Fig.~\ref{Fig:Fig2}A. At each
momentum $\mathbf{k}$, we compare the experimentally reconstructed density
matrix $\rho _{\mathbf{k}}$ with the ideal ground state $\left\vert \Psi _{%
\mathbf{k}}\right\rangle $\ and calculate the state fidelity $F_{\mathbf{k}%
}=\left\langle \Psi _{\mathbf{k}}\right\vert \rho _{\mathbf{k}}\left\vert
\Psi _{\mathbf{k}}\right\rangle $. The measured fidelities for different
momenta $\mathbf{k}$ are shown in Fig.~\ref{Fig:Fig2}B. We have achieved a
high average fidelity of $99.2\%$ in our experiment. A large fraction of the $0.8\%$ infidelity
is actually from the statistical error associated with a finite number of photon-counts
in the spin detection, which contributes to about $0.6\%$ infidelity (see the Supplementary Information). From the measured data,
we find the Hopf index shown in Fig.~\ref{Fig:Fig2}C for two different
phases with $h=0,2$. The measured non-zero values of the Hopf index, close
to the ideal integer numbers, provide an unambiguous experimental signature
for the underlying topological structure of the Hopf insulator phase.

Different from stacking layers of 2D quantum Hall insulators, the Hopf
insulator is an intrinsic 3D topological insulator, where the Chern numbers
characterizing 2D topological insulators are zero for all 2D momentum layers~%
\cite{moore2008topological, Deng2013Hopf}. To demonstrate this intuitively,
in Fig.~\ref{Fig:Fig3}A we take a layer in the momentum space (e.g., with $%
k_{z}=0$) and map all the measured spin states at different $k_{x}$ and $%
k_{y}$ to the Bloch sphere. The Chern number will be zero if these states
cannot fully cover the Bloch sphere, which is the case in Fig.~\ref{Fig:Fig3}%
a. We have also computed the Chern numbers explicitly using our measured
data along different 2D momentum layers \cite{Deng2014Direct} and checked
they are all identically zero. On the other hand, if we map the spin states
at all 3D momentum points to the Bloch sphere, they fully cover the sphere
as shown in Fig.~\ref{Fig:Fig3}B. This provides an intuitive indicator that
the Hopf insulator is an intrinsic 3D topological insulator.

\subsection{Observation of topological links}

\begin{figure}[t]
\includegraphics[trim=0cm 0cm 0cm 0cm, clip,width=\linewidth]{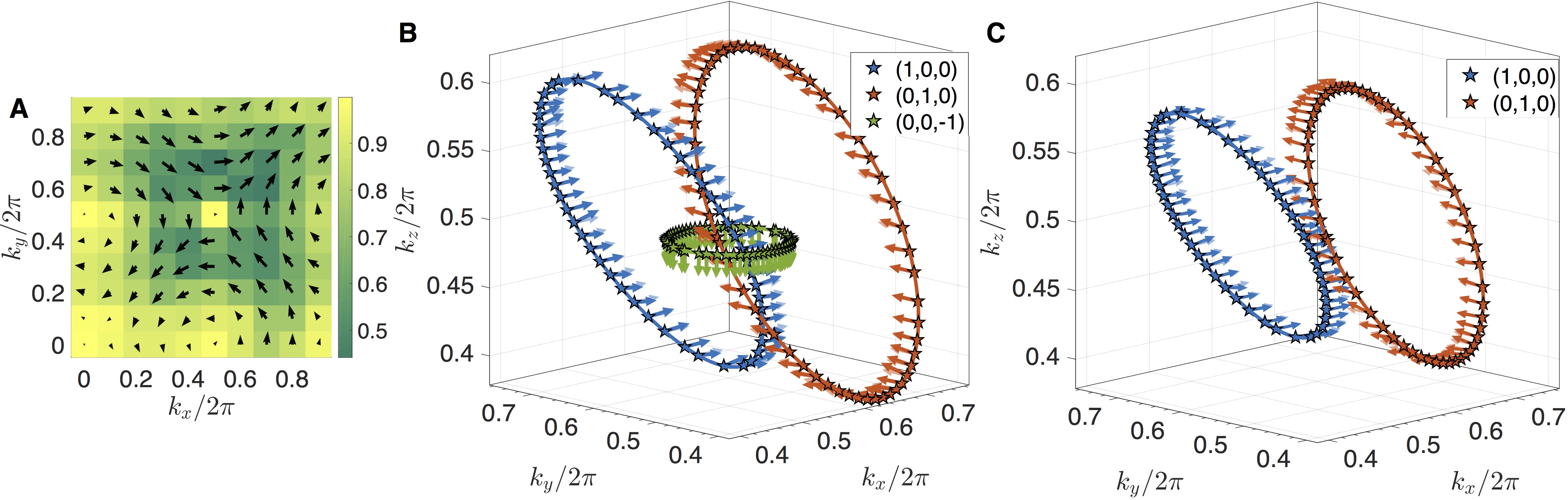}
\caption{\textbf{The Hopfion spin texture and topological links
characterizing the Hopf insulator.} (\textbf{A}) A 2D cross section of the
Hopfion is shown for the $k_{z}=0$ layer and $h=2$. The Hopfion represents
the knotted 3D spin texture of the Hopf insulator. The arrows in the plane
depict the $x$-$y$ direction of the Bloch vectors and the color labels the
magnitude of its $z$ component. (\textbf{B}) Topological links between the
pre-images in the momentum space from three spin states on the Bloch sphere,
$\mathbf{S} = \left( 1,0,0\right) $, $\left( 0,1,0\right) $, and $\left(
0,0,-1\right)$. The parameter $h=2.9$, which determines the topological
phase with the Hopf index $\protect\chi=1$. The linking number between any
pair of pre-image loops is equal to the Hopf index for this phase. The Solid
arrows show experimentally measured spin orientations $\mathbf{S}_{\text{exp}%
}$, which are in close proximity to the transparent arrows corresponding to
the ideal theoretical directions $\mathbf{S}_{\text{th}}$. Solid lines are
preimage curves from theoretical calculations. We find the average deviation
$\overline{|\mathbf{S}_{\text{exp}}-\mathbf{S}_{\text{th}}|} \approx 0.082,
0.076, 0.063$ for the blue, red and green curves. (\textbf{C}) When we cross
the topological phase transition point at $h=3$ and move to $h=3.1$, the
pre-image loops become unlinked, which implies that our system is in the
trivial phase at $h=3.1$. We have $\overline{|\mathbf{S}_{\text{exp}}-%
\mathbf{S}_{\text{th}}|} \approx 0.082, 0.071$ for the blue and red curves.
The preimage loop for the spin state $(0,0,-1)$ shrinks and vanishes at $%
h=3.1$ because none of the ground state of $H_{\mathbf{k}}$ could reach the
Bloch vector $(0,0,-1)$. This is another indication that the system is in
the trivial phase at $h=3.1$, since the ground-state spin orientations do
not fully cover the Bloch sphere.}
\label{Fig:Fig4}
\end{figure}

In the Hopf insulator, the momentum-space spin texture forms a
representation of the sought-after knotted structure known as the Hopfion
\cite{Faddeev1997Stable}. Distinct from 2D Skyrmion spin configurations
where swirling orientations are a salient feature \cite{Yu2010Real}, a 3D
Hopfion exhibits a nontrivial twisting. In Fig.~\ref{Fig:Fig4}A, we show a
cross-section of the measured spin texture along the $k_{z}=0$ layer, which
provides a glimpse of the 3D twisting of the Hopfion (see Supplementary
Information for the full 3D spin texture). If we fix a spin orientation on
the Bloch sphere and trace its pre-image in the momentum space $\mathbb{T}%
^{3}$, a closed loop will be formed. For the Hopf insulator phase, the loops
corresponding to different points (spin states) on the Bloch sphere are
always topologically linked to each other (see Hopf fibration in Fig.~\ref%
{Fig:Fig1}A). In order to measure these topological links, we experimentally
fix three spin states on the Bloch sphere and scan the momentum $\mathbf{k}$
so that the ground state of $H_{\mathbf{k}}$ is along these spin directions.
In Fig.~\ref{Fig:Fig4}B, we show the pre-images in the momentum space for
the spin states $\mathbf{S} = \left( 1,0,0\right) $, $\left( 0,1,0\right) $,
and $\left( 0,0,-1\right) $ on the Bloch sphere with the parameter $h=2.9$.
Clearly, these pre-images each form a closed loop and they are pairwise
linked with a linking number exactly equal to the Hopf index $\chi$. When we
cross the topological phase transition point (at $h=3$) and enter the
topologically trivial phase (with $h=3.1$), the corresponding pre-images are
shown in Fig.~\ref{Fig:Fig4}C---they are no longer linked. Hence the Hopf
links observed here provide a direct and intuitive characterization of
topological properties of the Hopf insulator.

\section{Discussion}

Our experimental probe of Hopf insulators in a solid state quantum simulator
represents the first experimental observation of 3D topological insulators
with integer-valued topological invariants and it paves the way for study of
complicated topological models using the powerful quantum simulation toolbox
\cite{Cirac2012Goals, Georgescu2014Quantum, Roushan2014Observation,
Kim2010Quantum}. The framework directly carries over to other types of
topological models that are predicted to exist in the periodic table \cite%
{Schnyder:2008ez, Kitaev2009Periodic, Deng2014Systematic} but have not yet
been observed in any experiments. The observation of Hopf links in our
experiment reveals the fascinating topological structure of the Hopf
fibration, and similar techniques can be employed to study higher index Hopf
insulators \cite{Deng2013Hopf} that can host a range of complex knots and
links \cite{Deng2016Probe}. This raises the possibility for experimental
exploration of the intimate connection between topological insulators and
mathematical knot theory. Based on measurement of momentum space
correlations, our first detection of an integer-valued topological invariant
in 3D complements well with other methods for measuring topological
invariants in 1D \cite{Atala2013Direct, Kong2016Direct} and 2D \cite%
{Roushan2014Observation, Schroer2014Measuring, Aidelsburger2015Measuring,
Mittal2016Measurement, Goldman2016Topological} systems. The demonstrated detection
scheme is general and applicable to probe of different topological
insulators in any spatial dimension predicted by the periodic table.


%

\textbf{Acknowledgements} This work was supported by the grants from the Ministry of
Science and Technology of China and the Ministry of
Education. L.M.D. and S.T.W. acknowledge in
addition support from the ARL and the AFOSR MURI programs. D.L.D. is
supported by JQI-NSF-PFC and LPS-MPO-CMTC.

\textbf{Author Contributions} L.M.D. conceived the experiment and supervised
the project. S.T,W. and D.L.D. performed theoretical analysis. X.X.Y., L.H.,
F.W., W.Q.L., X.W., C.H.Z., H.L.Z., X.Y.C. carried out the experiment. L.M.D.,
S.T.W. and X.X.Y. wrote the manuscript.

\textbf{Author Information} Correspondence and requests for materials should
be addressed to L.M.D. (lmduan@umich.edu).

\onecolumngrid

\newpage

\section{Supplementary Materials: Observation of topological links associated with Hopf insulators in a solid-state quantum simulator}

\section{Materials and Methods}

\subsection{Experimental setup}

We illuminate the diamond Nitrogen-Vacancy (NV) center and collect its fluorescence with a home-built confocal microscope with a $1.49$ NA objective (Olympus 60X, under oil immersion). A $532\,$nm diode laser (Coherent Sapphire), controlled by an acoustic-optical modulator (AOM,ISOMET 1250C) under double pass configuration, is used to initialize and detect the electron spin. Microwave from a signal generator (Keysight N5181B) is mixed with an arbitrary waveform generator (AWG, Tek AWG70002A) by an IQ mixer (Marki IQ 1545LMP) for phase and amplitude control. A waveguide with impedance matching is deposited onto the cover glass to deliver the microwave. The sample is mounted on the cover glass, which is then mounted on a closed-loop piezo (Physik Instrument P-611.3) with sub-micron resolution.

The sample is a type IIa single crystal diamond synthesized by chemical vapor deposition (Element Six). It is irradiated by $10\,$MeV electron beam with dosage $10^{14}\,$cm$^{-2}$ and annealed at $800^{\circ}$C in vacuum for two hours. Solid immersion lens with $10\, \mu$m radius is  fabricated on the sample by a focused ion beam. We get about $2.5\times 10^5$ counts per second in the single-photon detector with a signal-to-noise ratio $21 \colon 1$ through collection by a single mode fiber. A magnetic field of strength $510\,$G from a permanent magnet is applied along the NV axis of the sample to polarize the nuclear spins through spin flip-flop via the excited-state level anti-crossing \cite{Jacques2009Dynamic}.

\subsection{Adiabatic passage and data collection}

The final Hamiltonian $H_{\mathbf{k}} = \hbar
\Omega \, \mathbf{u}(\mathbf{k})\cdot \mathbf{\sigma }$ at a particular momentum point $\mathbf{k}$ is first normalized to \sloppy $\max \{ \Omega \sqrt{u_x^2+u_y^2}, \Omega u_z \}= 2\pi \times 20.83\,$MHz. For each adiabatic passage, we start from $u_{x} = u_{y} = 0$, $\Omega u_{z} = -2\pi \times 20.83\,$MHz under the initial state $\ket{0}$, which is the ground state of the initial Hamiltonian. We first linearly ramp up $\sqrt{u_x^2+u_y^2}$ to the maximum Rabi frequency $2\pi \times 20.83\,$MHz, then linearly ramp $u_{z}$ to the corresponding final value, and linearly ramp down  $u_{x}$ and $u_{y}$ to the final $\mathbf{u}(\mathbf{k})$ before taking measurements. Each linear ramp takes $500\,$ns, with the total time of $1.5\,\mu$s for the microwave control. In order to get high-fidelity data, the AWG works at $8\,$ GHz sampling rate.

At each momentum point $\mathbf{k}$, the initialization, adiabatic passage, and
measurements are repeated $1.25 \times10^6$ times, collecting about $9.3 \times 10^4$ photons. For each round, the fluorescence of the final state from the adiabatic passage is compared with the fluorescence under the $\ket{0}$ state, where the latter is used for normalization. Experimental density matrices are obtained by state tomography through the maximal likelihood estimation. The fidelity of each density matrix is calculated by comparing it with the ideal state,
and the fidelities from all the measured momentum points are distributed with median $99.59\%$ ($99.7\%$-$95.51\%$ with $95\% $ confidence interval) and mean 99.2\%.  A large contribution to the infidelity is actually from the statistical error associated with a finite number of photon counts. From the numerical simulation, we find that even with a perfectly prepared quantum state, the statistical error alone with the same number of photon counts as we collected in experiments will give a fidelity distribution with median $99.88\%$ ($100\%$-$97.26\%$ with $95\%$ confidence interval) and mean 99.4\%. So the fidelity of the prepared state by the process of adiabatic passage is high due to the existence of a significant energy gap for the topological phase.

\section{Measuring the Hopf invariant and theoretical scaling}

In this section, we present more details on our methods to extract the Hopf index from experimental data. We also theoretically extrapolate the data to larger grid sizes.
The central idea is to use tomographic methods to measure the topological invariant. This follows primarily from Ref.~\cite{Deng2014Direct, Deng2016Probe}. We outline the essential procedure here for completeness.

The Hopf invariant is defined in Eq.~(4) in the main text. In the experiment, we simulate the ground state of the Hopf Hamiltonian by adiabatically ramping from the $\ket{0}$ state (ground state of the Hopf Hamiltonian at $\mathbf{k} = (0, 0, 0)$) to other discrete momentum points $\mathbf{k_{J}}$. We can subsequently perform state tomography to map out the ground state manifold $\ket{\psi(\mathbf{k_{J}})}$. However, to calculate the Berry curvature from the states involves taking the derivatives $\partial_{k_{\nu,\tau}}$ (finite difference in our discrete data). This will lead to problems due to the gauge (phase) ambiguities of the wavefunction $\ket{\psi(\mathbf{k_{J}})} \to e^{i \varphi (\mathbf{k_{J}})} \ket{\psi(\mathbf{k_{J}})}$, where $e^{i \varphi (\mathbf{k_{J}})}$ is an arbitrary phase that can vary with $\mathbf{k_{J}}$ and is not experimentally observable. In cases where the Chern number is nonzero, gauge obstruction, in particular, forbids a well-defined global smooth Berry connection. To circumvent this difficulty, we use a discretized version of the Berry curvature defined as~\cite{Fukui2005Chern, Deng2014Direct}
\begin{align}
\label{Eq:BerryCurvature}
\mathcal{F}_{\mu}(\mathbf{k}_{\mathbf{J}}) & \equiv  \dfrac{i}{2\pi} \epsilon_{\mu\nu\tau} \ln U_{\nu} (\mathbf{k}_{\mathbf{J}})U_{\tau}(\mathbf{k}_{\mathbf{J+\hat{\nu}}}),
\end{align}
where the $U(1)$-link is $U_{\nu} (\mathbf{k}_{\mathbf{J}})\equiv\langle \psi(\mathbf{k}_{\mathbf{J}})|\psi(\mathbf{k_{\mathbf{J+\hat{\nu}}}})\rangle/|\langle \psi(\mathbf{k}_{\mathbf{J}})|\psi(\mathbf{k_{\mathbf{J+\hat{\nu}}}})\rangle|$ with $\hat{\nu}=\hat{x},\hat{y},\hat{z}$, a unit vector in the corresponding direction. Here, the local gauge ambiguity cancels out.

This tomographic method offers a number of advantages \cite{Deng2014Direct}. First, it is generally applicable to any spatial dimension and to all topological invariants that can be expressed as some variant of an integral over Berry curvature (connection). Second, the topological invariants can be extracted from the states alone, without referencing to the Hamiltonian.
Third, this method is highly robust to experimental imperfections and, in particular, finite discretizations.

In the experiment, we perform state tomography at various momentum points $\mathbf{k}_{\mathbf{J}}$. Discrete Berry curvature is then computed using Eq.~\eqref{Eq:BerryCurvature}. Berry connection $A_{\mu}(\mathbf{k}_{\mathbf{J}})$ can be obtained by Fourier transforming the equation $\nabla\times\mathbf{A}=\mathbf{F}$ with the Coulomb gauge $\nabla\cdot\mathbf{A}=0$. Finally, we attain the value of the Hopf invariant $\chi$ by a discrete sum over all momentum points. As we notice from the main text, a grid size of $10 \times 10 \times 10$ is already capable of producing highly accurate estimation of the quantized topological invariant (with error $\leq 5\%$).

\begin{figure}[p]
\includegraphics[trim=0cm 0cm 0cm 0cm, clip,width=\columnwidth]{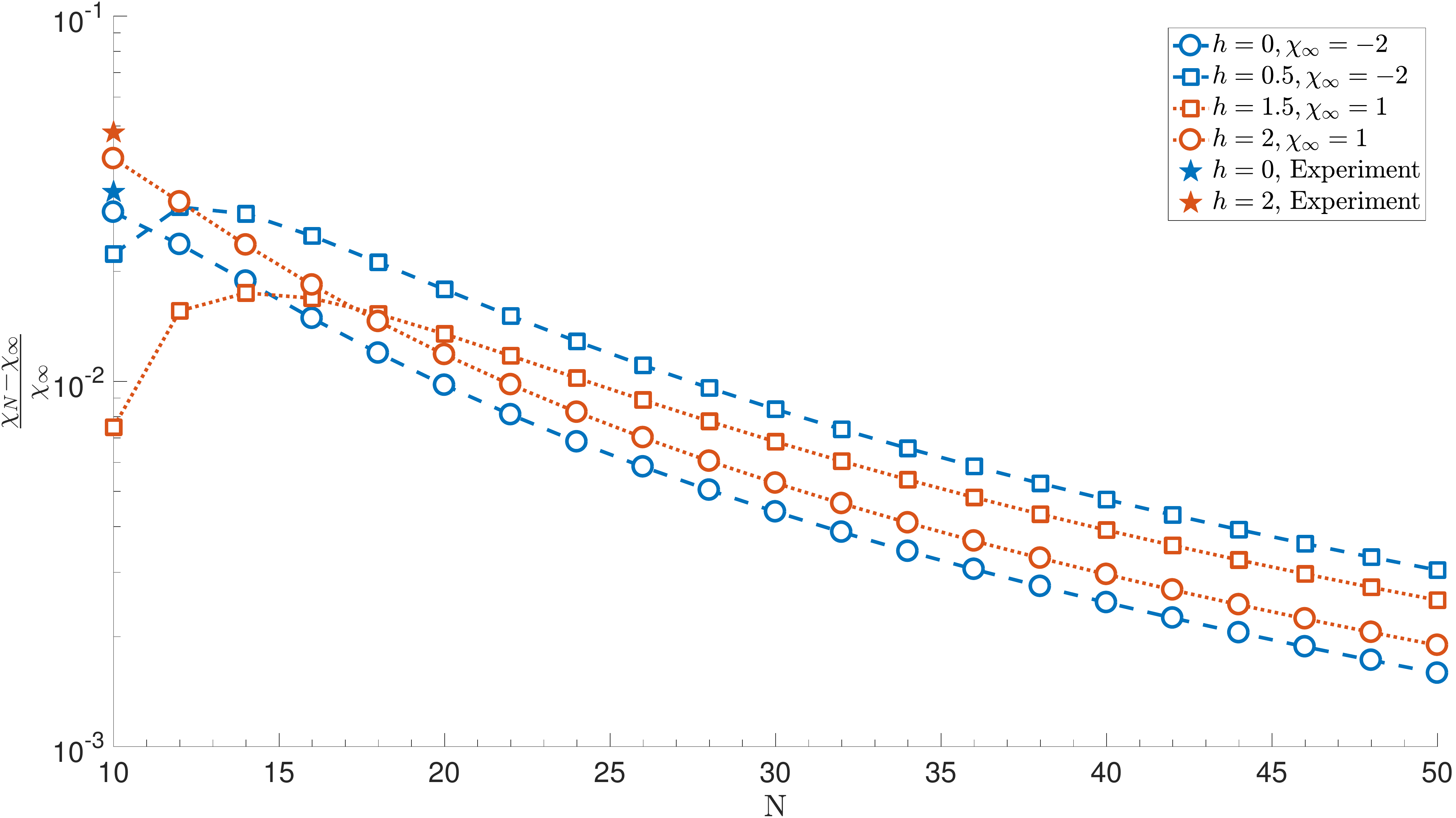}
\caption{Theoretical scaling of the deviation of the Hopf index $\chi_{N}$ from the ideal value $\chi_{\infty}$. The grid size is $N \times N \times N$. Experiments are performed at $N=10$. The apparently smaller deviation in the case of $h=0.5,1.5$ for $N \leq 15$ is likely to be coincidental.}
\label{Fig:HopfGridTh}
\end{figure}

In Fig.~\ref{Fig:HopfGridTh}, we present theoretical scalings to larger grid sizes. We can see that the discretization error reduces when $N$ increases. The deviation from the quantized value drops to around $10^{-2}$ for $N = 20$ for $h=0$ and $h=2$. The theoretical calculations for $h = 0.5$ and $h=1.5$ are also shown. They are closer to the topological phase transition point $h=1$, resulting in a more pronounced finite size effect. It is apparent, however, for all cases the finite-grid estimation approaches the correct quantized value as $N$ becomes larger. The topological property is robust to perturbations and changes in parameters as long as the topological gap is maintained.

\section{Three dimensional Hopf spin texture}

\begin{figure}[p]
\includegraphics[trim=0cm 0cm 0cm 0cm, clip,width=\columnwidth]{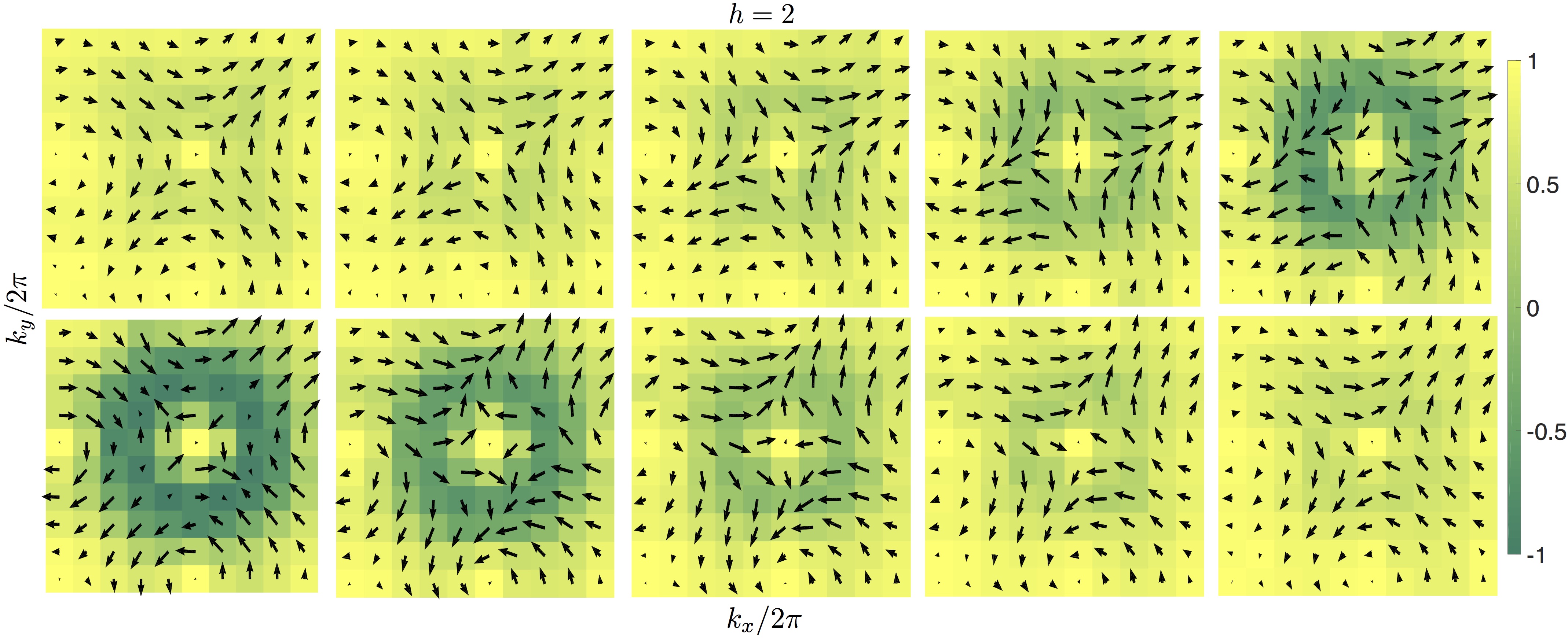}
\caption{Each layer of measured spin textures for $k_{z}=0,0.1,0.2, \cdots \!, 0.9 \times 2\pi$ and $h=2$. For each subfigure, $k_x/2\protect\pi$ and $k_y/2\protect\pi$ vary from  $0$ to $0.9$ with an equal spacing of $0.1$. At each momentum point $\mathbf{k_{J}}$, the state can be represented on the Bloch sphere. The arrows in the plane depict the $x$-$y$ direction of the Bloch vector and the color labels the magnitude of the $z$ component of the Bloch vector. This 3D spin texture represents a Hopfion with a Hopf invariant $\chi =1$.}
\label{Fig:spin2D_h2}
\end{figure}

\begin{figure}[p]
\includegraphics[trim=0cm 0cm 0cm 0cm, clip,width=\columnwidth]{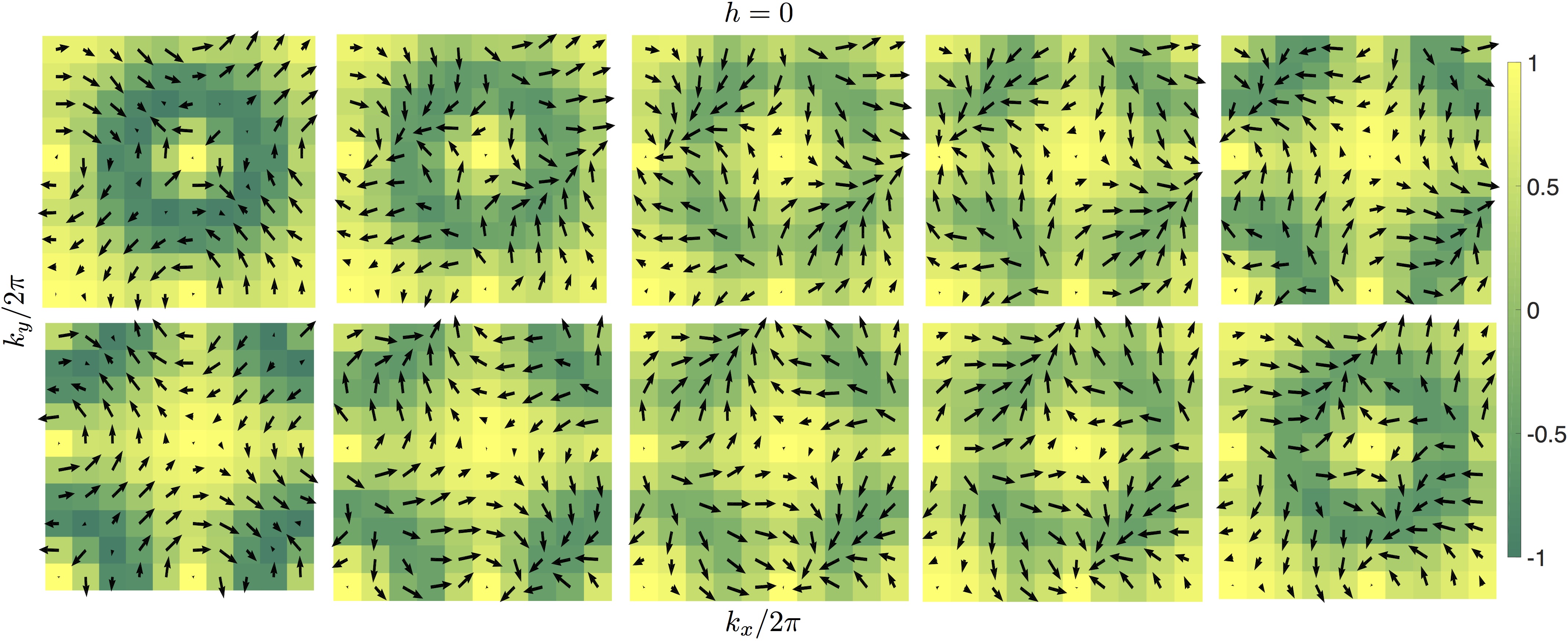}
\caption{Each layer of measured spin textures for $k_{z}=0,0.1,0.2, \cdots \!, 0.9 \times 2\pi$ and $h=0$. Spin representations and color scheme are the same as in Fig.~\ref{Fig:spin2D_h2}. This 3D spin texture represents a Hopfion with a Hopf invariant $\chi = -2$.}
\label{Fig:spin2D_h0}
\end{figure}

A two-dimentional (2D) slice of the spin texture is presented in the main text. Here, we include the full 3D spin texture from experimental data for both $h=2$ (Fig.~\ref{Fig:spin2D_h2}) and $h=0$  (Fig.~\ref{Fig:spin2D_h0}). Since the Hopf insulator is an intrinsic 3D topological insulator, complete information can be captured only by the 3D spin texture. For $h=0$, we have a higher (magnitude) topological index $\chi =-2$, so the spin texture is considerably more complex than that for $h=2$. Physically, a nonzero Hopf index guarantees the spin texture can never be untwisted to be a trivial one (e.g.\ all point to the same direction), unless one crosses a topological phase transition. Remarkably, the Hopf spin texture is a representation of the long sought-after Hopfions, which are 3D topological solitions with widespread applications~\cite{Deng2016Probe}.

\section{ Hopf Fibration \& Stereographic coordinates}
For simplicity and clarity, we did not use stereographic coordinates to represent the experimental data in the main text. The data were depicted in $\{ k_{x}, k_{y}, k_{z} \} \in [0, 2\pi)$ without gluing the boundaries. It does not matter for the particular spin preimage contours we measured because they form closed loops without crossing the boundaries (i.e., $k_{x,y,z}=0$ or $2\pi$). However, it may not be the case for other spin preimages, especially for higher Hopf index. When that happens, we have to visualize it properly on the torus $\mathbb{T}^{3}$; however, knots and links on the torus are difficult to see. Instead, we can map them to $\mathbb{R}^{3}$ for visualization. Indeed, our Hamiltonian mapping from the Brillouin zone $\mathbb{T}^{3}$ to the Bloch sphere $\mathbb{S}^{2}$ can be decomposed to two maps~\cite{Deng2016Probe}
\begin{equation}
\mathbb{T}^{3} \overset{g}{\longrightarrow} \mathbb{S}^{3} \overset{f}{\longrightarrow}  \mathbb{S}^{2}.
\end{equation}
The map $g$ is
\begin{align}
\eta_{\uparrow}\mathbf{(k)} & = \sin k_{x}-i\sin k_{y}, \notag \\
\eta_{\downarrow}\mathbf{(k)}& = \sin k_{z}-i(\cos k_{x}+\cos k_{y}+\cos k_{z}+h), \label{eq:map_g}
\end{align}
where $(k_{x}, k_{y}, k_{z})$ lives on $\mathbb{T}^{3}$ and $(\eta_{1},\eta_{2},\eta_{3},\eta_{4}) =(\text{Re}[\eta_{\uparrow}],\text{Im}[\eta_{\uparrow}],\text{Re}[\eta_{\downarrow}],\text{Im}[\eta_{\downarrow}])$ are points on $\mathbb{S}^{3}$ (up to a trivial normalization). The map $f$ is the Hopf map
\begin{equation}
u_{x}+i u_{y}=2\eta_{\uparrow}\bar{\eta}_{\downarrow},\; u_{z}= (|\eta_{\uparrow}|^{2}-|\eta_{\downarrow}|^{2}), \label{eq:HopfMap}
\end{equation}
and the composition of the two maps produces the Hamiltonian written in the main text, $H_{\mathbf{k}}/\hbar \Omega = f \circ g (\mathbf{k}) = \mathbf{u}(\mathbf{k})  \cdot \sigma$. Therefore, the knots and links can be visualized in $\mathbb{R}^{3}$ from the stereographic coordinates of $\mathbb{S}^{3}$, for example, defined as
\begin{equation}
(x,y,z)  =  \frac{1}{1+\eta_{4}}(\eta_{1},\eta_{2},\eta_{3}),
\end{equation}
where $(x,y,z)$ are points of $\mathbb{R}^{3}$.
In Fig.~1A of the main text, the Hopf fibration is drawn under the Hopf map $f$. For a fixed point on the Bloch sphere $\mathbb{S}^{2}$, the preimage (fiber) of the point forms a closed loop in $\mathbb{S}^{3}$, which is then visualized in $\mathbb{R}^{3}$ via the stereographic coordinates. To relate the schematic to our physical system, the preimage of a fixed spin orientation measurement (on the Bloch sphere) lives in the momentum space $\mathbb{T}^{3}$, which can then be mapped to $\mathbb{S}^{3}$ via the map $g$ and subsequently $\mathbb{R}^{3}$ via the stereographic coordinates.

The Hopf map from $\mathbb{S}^{3} \to \mathbb{S}^{2}$ can be modified to the generalized Hopf map \cite{Deng2013Hopf}
where a variety of knot and link structures can be revealed from Hopf insulators \cite{Deng2016Probe}. We emphasize that the change of coordinates is only for the purpose of easy visualization. The nontrivial link induced by the nonzero Hopf invariant cannot be unlinked, in either $\mathbb{R}^{3}$, $\mathbb{S}^{3}$ or $\mathbb{T}^{3}$ since the maps between them are all continuous.

\section{Finite resolution \& $\epsilon$-neighborhood of spin orientations}

To reveal the nontrivial Hopf fibration and linking structures of the spin preimage loops, in the main text, we took experimental data on the theoretically known contours. We observed that the experimentally measured spin orientations agree well with the theory (with fidelity $F \gtrsim 99\%$ and $|\mathbf{S}_{\text{exp}}-\mathbf{S}_{\text{th}}| \lesssim 0.08$). The nontrivial links as well as the topological phase transition were readily detected from experimental data. In situations where the theoretical contours are unknown, one has to measure the spin orientations at discrete momentum data grids and deduce the preimage loops with a prescribed tolerance threshold. In this case, we can define an $\epsilon$-neighborhood of the desired spin orientation, $\mathbf{S}_{\text{th}}$, as \cite{Deng2016Probe}
\begin{equation}
N_{\epsilon}(\mathbf{S}_{\text{th}})  =  \{ \mathbf{S}_{\text{exp}}(\mathbf{k}) : \; |\mathbf{S}_{\text{exp}}(\mathbf{k}) -\mathbf{S}_{\text{th}}|\leq\epsilon\}.
\end{equation}
The choice of $\epsilon$ depends on the actual experimental data; it should be chosen large enough to contain sufficient data points and small enough to display a clear loop structure. This scheme is also applicable to the case when we are presented with a 3D spin texture data and aim to ascertain whether it exhibits nontrivial knot or loop structures.

\begin{figure}[p]
\includegraphics[trim=0cm 0cm 0cm 0cm, clip,width=\columnwidth]{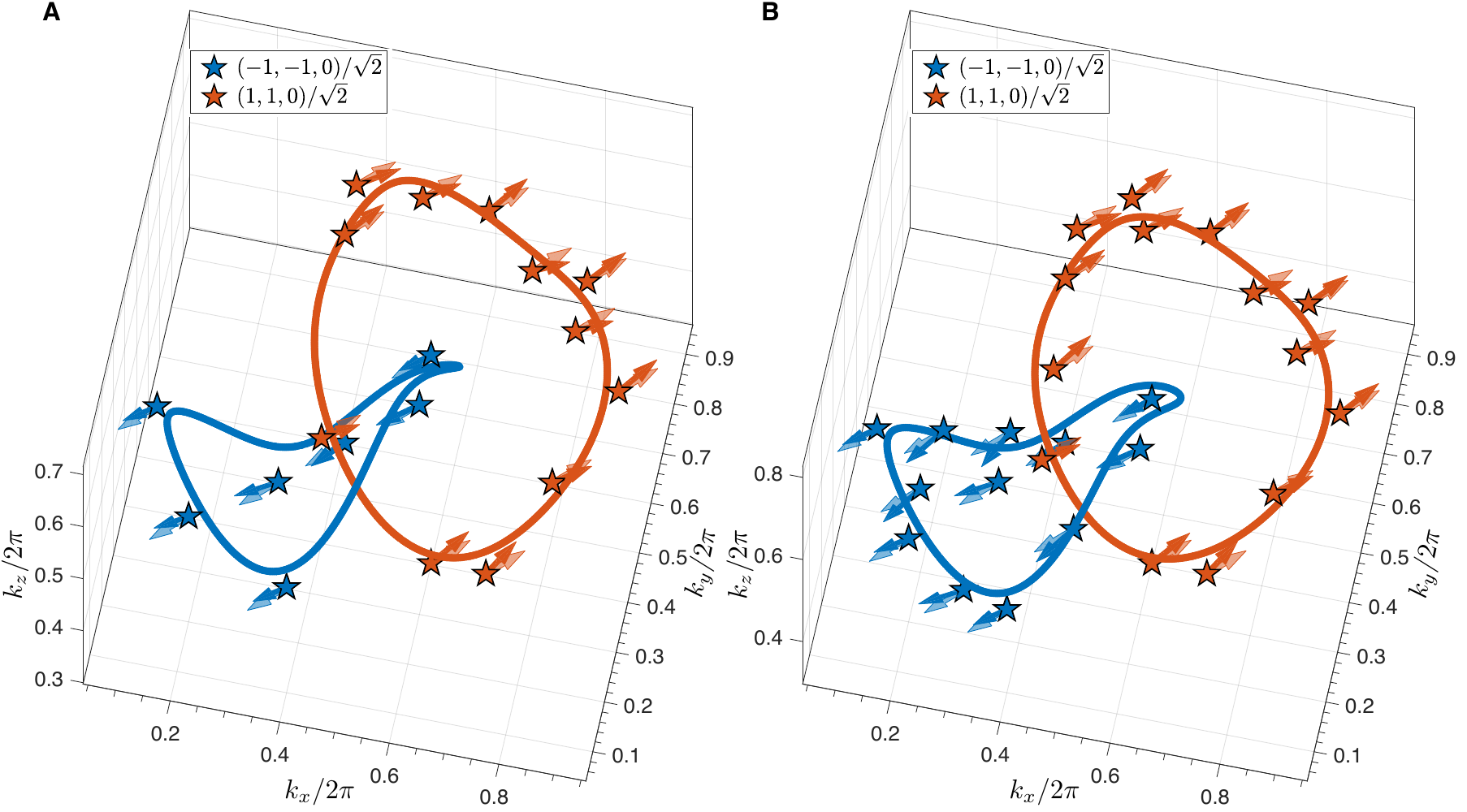}
\caption{Preimage contours using the experimental data on the $10 \times 10 \times 10$ grid for $h=2$. Stars are experimental data on the discrete grid satisfying the condition $|\mathbf{S}_{\text{exp}}-\mathbf{S}_{\text{th}}|\leq\epsilon$, where $\mathbf{S}_{\text{th}} = (-1,-1,0)/\sqrt{2} $ or $(1,1,0)/\sqrt{2}$ for the blue and red data respectively. Solid arrows show experimentally measured spin orientations, $\mathbf{S}_{\text{exp}}$. Transparent arrows show theoretical directions, $\mathbf{S}_{\text{th}}$, imposed on the experimental grids. Solid lines are theoretical preimage curves. (\textbf{A}) $\epsilon = 0.3$. (\textbf{B}) $\epsilon = 0.35$.
}
\label{Fig:SMHopfLink}
\end{figure}

To show the method works well with limited experimental data resolution and is reasonably robust to the choice of $\epsilon$, here we use our experimental data on the $10 \times 10 \times 10$ grid to map out the nontrivial loops at $h=2$. Fig.~\ref{Fig:SMHopfLink} shows the results with $\epsilon = 0.3$ and $\epsilon = 0.35$ respectively. With larger tolerance, it is evident that more experimental data points are included. Being a topological property, the nontrivial loop structure is reasonably robust to the choice of $\epsilon$. Imposing the theoretical curves as guides to the eye, the nontrivial link is discernible with experimental data. We remark that the discrepancies are predominantly due to the coarse discretization. With more experimental data on a finer grid, e.g.\ a $20 \times 20 \times 20$ grid, the preimage loops and the nontrivial links should be clearly visible even without the theoretical curve; they are also expected to be highly robust to small perturbations such as experimental errors, change in Hamiltonian parameters, and the choice of the tolerance threshold.

%

\clearpage

\clearpage

\clearpage

\clearpage

\end{document}